\documentclass[a4paper]{spie}  

\usepackage{graphicx,amsmath,cite}

\renewcommand{\l}{\left}
\renewcommand{\r}{\right}

\newcommand{\sign}{{\rm sign}}
\newcommand{\erf}{{\rm erf}}
\newcommand{\avg}[1]{\left\langle{#1}\right\rangle}
\newcommand{\ovl}[1]{\overline{#1}}
\newcommand{\ii}{{\rm i}}
\newcommand{\cavg}[1]{\left\langle\!\left\langle{#1}
        \right\rangle\!\right\rangle}
\newcommand{\be}{\begin{equation}}
\newcommand{\ee}{\end{equation}}
\newcommand{\beg}{\begin{gather}}
\newcommand{\eeg}{\end{gather}}
\newcommand{\bsy}[1]{\boldsymbol{#1}}

\title{On the interplay between fluctuations and efficiency in a model economy with heterogeneous adaptive consumers}

\author{Andrea De Martino\supit{a} and Matteo Marsili\supit{b}
\skiplinehalf
\supit{a}INFM SMC, Dipartimento di Fisica, Universit\`a di Roma ``La Sapienza''\\p.le A. Moro 2, 00185 Roma (Italy)\\
\supit{b}The Abdus Salam ICTP, Strada Costiera 11, 34014 Trieste (Italy)
}

\authorinfo{E-mail: andrea.demartino@roma1.infn.it, marsili@ictp.trieste.it}

\begin{document}
\maketitle 

\begin{abstract}
We discuss the stationary states of a model economy in which $N$ heterogeneous adaptive consumers purchase commodity bundles repeatedly from $P$ sellers. The system undergoes a transition from an inefficient to an efficient state as the number of consumers increases. In the latter phase, however, price fluctuations may be much larger than in the inefficient regime. Results from dynamical mean-field theory obtained for $N\to\infty$ compare fairly well with computer simulations.
\end{abstract}


\keywords{SPIE Proceedings: Fluctuations and noise 2005 -- Austin, TX}

\section{Introduction}

Systems of heterogeneous units interacting competitively -- like producers and consumers in modern economies, traders in financial markets, drivers on city roads and highways or users accessing a computer network -- pose serious coordination problems of both theoretical and practical relevance. It is reasonable to think that each agent in such situations aims at optimizing his or her individual performance making decisions on the basis of available information and experience. However, private goals will most certainly be conflicting, which makes the reach of a globally efficient phase where the collective use of resources is optimal far from certain. More generally, the interplay between different macroscopic properties in these systems may turn out to be quite subtle. It is clear that one of the crucial points that a theory of these systems should address is what resource load distributions may emerge and under which conditions. Unfortunately, the mathematical framework of general equilibrium theory \cite{mas} does not appear to be able to provide answers to these problems. Its main stumbling block consists in the fact that, from a technical viewpoint, it is rather difficult to extract robust macroeconomic laws from the underlying microeconomic assumptions while maintaining the crucial ingredient of heterogeneity \cite{kirman} (see also \cite{foll,noi1} for alternative approaches that try to overcome these problems). 

On the other hand, agent-based models in which agents' behavior is described by simplified stochastic laws are often more amenable to tackle these issues. A remarkable example is given by the Minority Game \cite{book,cool}, whose elementary framework allows to elucidate many important aspects by addressing directly the relation between microscopic behavior and macroscopic properties (like fluctuations, predictability and efficiency). From a physical viewpoint, these models are relatives of mean-field spin glasses \cite{mpv} and neural networks \cite{hkp}. They can thus be studied in detail using the statistical mechanics toolbox for disordered systems, by which one can derive equations for the relevant observables while fully preserving heterogeneity at the level of agents.

Here we introduce and study both analytically and numerically a model closely related to both the batch Minority Game \cite{hc} and to the traffic model introduced in \cite{traf} in which the subtle interdependence of macroscopic properties is particularly striking. We aim at describing the adaptive dynamics of $N$ consumers who on each day select their consumption among $P$ possible commodities or producers using a minimum expected cost criterion and learning from experience. The system turns out to reach a steady state whose macroscopic properties, such as the distribution of consumer choices over commodities, are to a large extent independent of the microscopic details for $N\to\infty$. Upon increasing the ratio $N/P$ one observes a transition from a regime where some commodities are over- or under-used (and therefore some sellers are more convenient than others) and the average price is high to one in which consumers are uniformly distributed over resources and the average price is low. However while the former phase is characterized by contained price fluctuations, in the latter fluctuations may be much larger. This type of scenario is not dissimilar to that found in Minority Games (although details are somewhat different). The present model however provides hopefully a more concrete ground for testing it against empirical data.

\section{Model definitions}

We consider a system with $N$ consumers and $P$ producers or sellers. For our purposes, one may either think that different producers sell different goods, or that each seller supplies a different variety of the same broad category of commodities (for instance, perishable goods). On each day $t=1,2,\ldots$, every consumer $i$ has to acquire one of $S$ possible bundles of commodities, for instance for his or her subsistence. A bundle is a vector $\bsy{q}_{ig}=\{q_{ig}^\mu\}$ such that $q_{ig}^\mu$ denotes the amount of goods $i$ demands from seller $\mu$ ($\mu\in\{1,\ldots,P\}$). $g\in\{1,\ldots,S\}$ labels the different bundles of each consumer. Consumers are heterogeneous, in the sense that different consumers have different needs and thus different possible bundles. We assume that sellers set the daily price of commodities according to the demand they receive, denoted by $D^\mu(t)$, so that the higher the demand the higher the price. Hence prices do not enter directly in the model, but just through the demands. Each consumer on the other hand aims at purchasing, on each day, the bundle he or she finds more convenient, labeled by $g_i(t)$, with the limitation that when the choice is made the price at which the purchase will take place is not known yet (it is determined by the collective decision of all consumers, which form the demands). Hence they try to learn the convenience of different bundles from experience in order to be able to predict which bundle will have the highest marginal utility on any given day. The events taking place on each day $t$ can be summarized by the following scheme:
\begin{gather}
g_i(t)={\rm arg }\max_g U_{ig}(t)\label{decision}\\
D^\mu(t)=\frac{1}{N}\sum_i q_{i g_i(t)}^\mu\label{aggregate}\\
U_{ig}(t+1)-U_{ig}(t)=-\frac{1}{P}\sum_\mu q_{ig}^\mu\l[D^\mu(t)-k\r]\label{updating}
\end{gather}
At the decision stage, Eq. (\ref{decision}), each consumer chooses the bundle which carries the highest (cumulated) utility $U_{ig}(t)$. The different choices are then aggregated, Eq. (\ref{aggregate}), and the (normalized) demands are formed. Finally, (\ref{updating}), utilities are updated with the following rationale: if the demand of a commodity $\mu$ is above a certain threshold $k$, consumers perceive that commodity as too costly and the utility of bundles including it will tend to be reduced; similarly, if the demand has been lower than $k$ the commodity will be seen as `cheap'  and will tend to increase the utility of the bundle. (This mechanism is conceptually identical to that employed in standard Minority Games.) The utility of the bundle is then determined by the demands of all commodities in the bundle though a simple average. (This is what makes the model more similar to a batch Minority Game.)

In principle, $k$ could be agent- and commodity-dependent. For the sake of conceptual and technical simplicity, we ignore this possibility here. We assume that bundles $\bsy{q}_{ig}$ are quenched random vectors with probability distribution
\be
P(\bsy{q}_{ig})=\prod_\mu\l[(1-q)\delta(1-q_{ig}^\mu)+q\delta(q_{ig}^\mu)\r]
\ee
($0<q<1$ being the probability that any given commodity is part of a bundle) that are assigned to consumers independently on $i$ and $g$ on day $n=0$ and are kept fixed. In this way, we introduce a further simplification in that each seller is either visited or not by a consumer, and the purchased quantities play no role. Moreover, we are implicitly assuming that the different goods are equivalent to consumers, that is there is no commodity that all consumers will need to buy. Finally, we assume that the learning dynamics (\ref{updating}) is initialized at values $U_{ig}(0)$ about which more will be said later on.

We concentrate our attention on the macroscopic properties of the steady state(s). The observable by which they will be characterized is given by the magnitude of demand fluctuations,
\be
\Delta=\frac{1}{P}\sum_\mu\l[\avg{(D^\mu)^2}-\avg{D^\mu}^2\r]
\ee
(here and in what follows, $\avg{\ldots}$ stands for a time average in the stationary state of (\ref{updating})). Because of our assumptions on the relation between prices and demands, $\Delta$ quantifies the typical spread of prices in the economy. A measure of how evenly consumers are distributed over producers is instead given by
\be
H=\frac{1}{P}\sum_\mu\avg{D^\mu-\ovl{D}}^2
\ee
where $\ovl{D}=1-q$ is the expected demand. If $H=0$, each seller receives on average the same demand so that none of them is perceived as more convenient. In this case, consumers are distributed uniformly over producers. If $H>0$, instead, the distribution of demands is not uniform and some producers are seen as more or less convenient than others. When $H>0$ an external agent who watches the economy from the outside trying to identify the best bargain would manage to find more convenient sellers and make a profit. When $H=0$, instead, this would not be possible. So one sees that transitions from regimes with $H>0$ to regimes with $H=0$ can be seen as transitions between inefficient and efficient states of the economy, where by efficient state we mean one where goods flow from producers to consumers in such a way that no information exploitable by an external agent is generated. States that are optimal from a collective perspective have both $H=0$ and $\Delta$ small, because on one hand efficiency is desirable and on the other price fluctuations should be such that agents have as much cost certainty as possible on a day by day basis. Hence $H$ and $\Delta$ describe intertwined properties, and it is on their mutual dependence that we shall focus in what follows.

\section{Dynamical solution for $S=2$ (sketch)}

The above model can be solved exactly for $S=2$ (two possible bundles per consumer) resorting to dynamical techniques developed in the context of mean-field spin glass theories, which allow to obtain a complete macroscopic characterization of the stationary states in the limit $N\to\infty$. In this limit, the most remarkable phenomenology is obtained when the number of sellers scales linearly with $N$, i.e. when $\lim_{N\to\infty}N/P=n$ is finite. This is the case we will consider henceforth. In this section we will limit ourselves to a sketch of the broad lines of this approach, which requires a calculation that can be carried out (modulo some necessary modifications) following the lines traced in \cite{hc} and subsequent papers for the batch Minority Game (see \cite{tobias} for the most recent work and for references). The resulting theory correctly describes the steady states for the particular choice $k=1-q$, that is for the case in which the threshold equals the expected price of commodities. This simplifies the calculation considerably. Different choices of $k$ (which may lead to significantly different physics) will not be considered here.

The solution relies on the introduction of the auxiliary variables
\be\label{defi}
\bsy{\xi}_i=\frac{\bsy{q}_{i1}-\bsy{q}_{i2}}{2}~~~~~~~
\bsy{\omega}_i=\frac{\bsy{q}_{i1}+\bsy{q}_{i2}}{2}~~~~~~~\text{and}~~~~~~~
p_i(t)=\frac{U_{i1}(t)-U_{i2}(t)}{2}
\ee
in terms of which the bundle selected by consumer $i$ on day $t$ can be written as $\bsy{q}_{ig_i(t)}= \bsy{\omega}_i+s_i(t)\bsy{\xi}_i$ with $s_i(t)=\sign[p_i(t)]$. The latter -- Ising spins -- are ultimately the relevant microscopic dynamical variables of the problem. We call $p_i$ the `preference' of agent $i$, since if $p_i(t)>0$ (resp. $p_i(t)<0$) he/she selects bundle $1$ (resp. $2$). The time evolution of the preferences is governed by the equation
\begin{equation}\label{proc}
p_i(t+1)-p_i(t) =  -\frac{1}{P}\sum_\mu\xi_i^\mu\l[\frac{1}{N}\sum_j 
\l[\omega_j^\mu+s_j(t)\xi_j^\mu\r]-k\r]+h_i(t)
\end{equation}
where we added a small external probing field $h_i(t)$ for later use. One sees that our system is described by a set of $N$ coupled Markovian `zero temperature' processes with quenched disorder. In this case, like often in models with mean-field interaction \cite{bckm}, the steady states can be completely described in the limit $N\to\infty$ in terms of two-time macroscopic correlation and response functions:
\be\label{funcs}
C(t,t')=\frac{1}{N}\sum_i \ovl{\cavg{s_i(t)s_i(t')}}~~~~~~~
G(t,t')=\frac{1}{N}\sum_i\frac{\partial\ovl{\cavg{s_i(t)}}}{\partial h_i(t')}
\ee
where the double brackets represent an average over realizations of (\ref{proc}) at fixed disorder $\{\bsy{\omega}_i,\bsy{\xi}_i\}$ and the over-line stands for an average over the disorder. The canonical method to obtain equations for these quantities in these problems consists in evaluating the dynamical partition function
\be
Z[\bsy{\psi}] =\ovl{ \cavg{e^{\ii\sum_{i,t} s_i(t)\psi_i(t)}}}=\int e^{\ii\sum_{i,t}s_i(t)\psi_i(t)} \rho(\bsy{p}(0)) ~\ovl{\prod_{i,t}\l[\delta\l({\rm equation~(\ref{proc})}\r)dp_i(t)\r]}
\ee
(with $\rho(\bsy{p}(0))$ the distribution of initial conditions) which satisfies
\be
C(t,t')=-\lim_{\{\psi_i\to 0\}}\frac{1}{N}\sum_i\frac{\partial^2 Z[\bsy{\psi}]}{\partial\psi_i(t)\psi_j(t')}~~~~~~~
G(t,t')=-\lim_{\{\psi_i\to 0\}}\frac{\ii}{N}\sum_i\frac{\partial^2 Z[\bsy{\psi}]}{\partial\psi_i(t)\theta_j(t')}
\ee
This can be done in two steps. First, the disorder average is carried out, generating two-time dynamical order parameters such as $Q(t,t')=(1/N)\sum_i s_i(t) s_i(t')$. Next, the limit $N\to\infty$ is performed via a saddle-point integration. At the relevant saddle, the dynamical order parameters can be identified with the correct correlation and response functions. This in turn leads to a process describing the behavior of a single `effective' consumer via a stochastic non-Markovian equation whose generic form is
\be\label{dmfe}
p(t+1)-p(t)=h(t)+\sum_{t'}R(t,t')s(t')+\eta(t)
\ee
where $R(t,t')$ is known as the retarded self-interaction kernel and $\eta(t)$ is a zero-average Gaussian noise with a non-trivial covariance matrix $\avg{\eta(t)\eta(t')}$. Different models ultimately result in different forms of $R(t,t')$ and of $\avg{\eta(t)\eta(t')}$. Both these quantities turn out to depend only on the correlation function $\bsy{C}=\{C(t,t')\}$ and on the response function $\bsy{G}=\{G(t,t')\}$, which can in turn be obtained from the statistics of the effective consumer's spin $s(t)=\sign[p(t)]$ generated by the dynamical mean-field equation (\ref{dmfe}):
\be
C(t,t')=\avg{s(t)s(t')}~~~~~~~G(t,t')=\frac{\partial\avg{s(t)}}{\partial h(t')}
\ee

For the model described here, the retarded self-interaction kernel reads
\be\label{rsi}
\bsy{R}=-\frac{q(1-q)}{n}\l[\bsy{1}+q(1-q)\bsy{G}\r]^{-1}
\ee
where $\bsy{1}=\{\delta_{tt'}\}$, while the noise is conveniently written as $\eta(t)=\sqrt{q(1-q)/n}~z(t)$ with covariance matrix $\avg{z(t)z(t')}=\Lambda(t,t')$ and
\be\label{noizcov}
\bsy{\Lambda}=\l[\bsy{1}+q(1-q)\bsy{G}\r]^{-1}\l[q(1-q)(\bsy{E}+\bsy{C})\r]
\l[\bsy{1}+q(1-q)\bsy{G}\r]^{-1}
\ee
where $\bsy{E}=\{1\}$ (for all $t,t'$). 

In principle, once (\ref{dmfe}) is obtained, it is possible to calculate $C(t,t')$ and $G(t,t')$ at all times (notice that time has been kept finite up to now). In what follows, we shall concentrate on extracting the macroscopic behavior in the limit $t\to\infty$. This will be later compared to numerical simulations.

\section{Stationary state of the dynamical mean-field equations: persistent order parameters and fluctuations}

In order to calculate the steady state properties, it is necessary to formulate suitable Ans\"atze for the asymptotic behavior of correlation and response functions. We assume here that the system reaches an ergodic steady state, in which both are time-translation invariant and there is no anomalous response. This corresponds to requiring that \cite{bckm}
\begin{gather}
\lim_{t\to\infty}C(t+\tau,t)=C(\tau)~~~~~~~\lim_{t\to\infty}G(t+\tau,t)=G(\tau)\label{tti}\\
\chi:=\lim_{\tau\to\infty}\sum_{t\leq \tau}G(t)<\infty\label{fir}\\
\lim_{t\to\infty}G(t,t')=0~~~~\forall t'{\rm ~finite}\label{wltm}
\end{gather}
(top to bottom: time translation-invariance, absence of anomalous response, absence of long-term memory). Numerically, one observes that in the steady state some agents always buy the same bundle while others keep flipping between their possible bundles. Borrowing Minority Game jargon, we call the former, for which $|p_i(t)|$ grows linearly with time, `frozen' and the latter, for which preferences remain finite as $t\to\infty$, `fickle'. Their respective contributions to the steady state properties can be studied separately. Defining $\widetilde{p}(t)=p(t)/t$ one has, from (\ref{dmfe}),
\be
\widetilde{p}(t)-\frac{1}{t}\widetilde{p}(1)=\frac{1}{t}\sum_{t'<t}\sum_{t''<t'}R(t',t'')s(t'')+\frac{1}{t}\sqrt{\frac{q(1-q)}{n}}
\sum_{t'<t}z(t')+h(t)
\ee
where $\bsy{R}$ is given by (\ref{rsi}) and the noise covariance by (\ref{noizcov}). Taking the limit $t\to\infty$ and neglecting the external field this becomes
\be
\widetilde{p}=-\frac{q(1-q)s}{n\l[1+q(1-q)\chi\r]}+\sqrt{\frac{q(1-q)}{n}} z
\ee
provided one defines
\be
\widetilde{p}=\lim_{t\to\infty}\widetilde{p}(t)~~~~~~~
s=\lim_{t\to\infty}\frac{1}{t}\sum_{t'}s(t')~~~~~~~
z=\lim_{t\to\infty}\frac{1}{t}\sum_{t'}z(t')
\ee
($\chi$ is the integrated response defined in (\ref{fir})). The variance of $z$ is obtained from (\ref{noizcov}):
\be
\avg{z^2}=\lim_{t,\tau\to\infty}\frac{1}{t\tau}\sum_{t'<t}\sum_{t''<\tau}\Lambda(t',t'')=\frac{q(1-q)(1+c)}{\l[1+q(1-q)\chi\r]^2}
\ee
where $c$ is the persistent autocorrelation $c=\lim_{t\to\infty}(1/t)\sum_{t'}C(t')$. Now for a frozen consumer $\widetilde{p}$ is non-zero and $s$ is either $1$ (for $\widetilde{p}>0$) or $-1$ (for $\widetilde{p}<0$). This leads to the condition $|z|>\gamma$ with 
\be
\gamma=\frac{\sqrt{q(1-q)/n}}{1+q(1-q)\chi}
\ee
Similarly, for a fickle consumer $\widetilde{p}=0$ and $s=z/\gamma$, which corresponds to $|z|<\gamma$. One may now easily derive a self-consistent equation for $c\equiv\avg{s^2}$:
\be\label{c}
c=\avg{\theta(|z|-\gamma)}+\avg{z^2\theta(\gamma-|z|)/\gamma^2}=
\phi+\frac{1}{\lambda^2}\l[\ovl{\phi}-\lambda\sqrt{\frac{2}{\pi}} e^{-\lambda^2/2}\r]
\ee
where $\phi=\avg{\theta(|z|-\gamma)}=1-\erf(\lambda/\sqrt{2})$ is the fraction of frozen consumers, $\ovl{\phi}=1-\phi$ is the fraction of fickle consumers, and $\lambda=\gamma/\sqrt{\avg{z^2}}=
1/\sqrt{n(1+c)}$. In order to obtain an equation for $\chi$, we can proceed as follows:
\be\label{chi}
\chi=\avg{\frac{\partial s}{\partial h}}=\frac{1}{\sqrt{q(1-q)/n}}\avg{\frac{\partial s}{\partial z}}
=\frac{\ovl{\phi}}{\gamma\sqrt{q(1-q)/n}}
\ee
where we used the fact that frozen agents are insensible to small perturbations and thus do not contribute to the response. Eq.s (\ref{c}) and (\ref{chi}) can be easily solved numerically to obtain $c$ and $\chi$ as a function of $n$ for different $q$. Notice that when $\chi$ diverges (\ref{fir}) is no longer valid and the current theory breaks down. Using the definition of $\gamma$, this turns out to happen when $n$ is such that $n\ovl{\phi}=1$. This condition is independent of $q$. Minor manipulations performed imposing it to (\ref{c}) lead to the critical value $n_c=2.9638\ldots$, which represents the point above which the ergodic solution obtained starting from (\ref{tti}), (\ref{fir}) and (\ref{wltm}) must be replaced by a non-ergodic one, in particular by one with long-term memory for which the steady state depends on the initial conditions of the dynamics. We will not consider this region analytically here. However, we will show by simulations that the above picture is correct.

As we stated before, most of our attention is placed, rather than on $c$ and $\chi$, on $\Delta$ and $H$. $\Delta$ is particularly hard to calculate exactly, since it is not in general a function of the persistent order parameters only (both long-term and short-term fluctuations contribute to it). However, it is possible to obtain rough estimates for both quantities in terms of $c$ and $\chi$ alone from the matrix $\bsy{\Xi}=\frac{1}{2}\bsy{\Lambda}$, using the approximate method first employed in \cite{hc} to calculate the volatility of the batch Minority Game. This procedure consists essentially in separating the persistent contribution to $\bsy{\Xi}$ from the non-persistent one and in neglecting the autocorrelation of fickle consumers. We shall omit details of the (long but straightforward) derivation here. Let it suffice to say that $H$ and $\Delta$ have the following approximations:
\begin{gather}
H=\frac{q(1-q)(1+c)}{2\l[1+q(1-q)\chi\r]^2}\\
\Delta=\frac{q(1-q)(\phi-c)}{2\l[1+q(1-q)\chi\r]^2}+\frac{1}{2}q(1-q)\ovl{\phi}
\end{gather}
Notice that at $n_c$, i.e. when $\chi$ diverges, $H$ vanishes. This means that $H$ behaves roughly as a physical order parameter: for $n>n_c$ the system reaches an ``ordered'' state in which the distribution of consumers over producers is uniform. The behavior in the limit $n\to 0$ can be obtained with a minimal algebraic effort. It turns out that in this limit $c\sim n$ so $\phi$ vanishes. The same can be said for $\chi$, so that finally $H\to q(1-q)/2$ and $\Delta\to q(1-q)/2$.

\section{Comparison with numerical results}

We have performed computer simulations of the model with $k=1-q$ for different $q$ and initial conditions, fixing the product $NP=16000$. We shall first discuss the case $S=2$, for which the theory outlined above holds. Later the dependence on $S$ will be addressed. We only consider the case $q\leq1/2$ since analytically all macroscopic properties turn out to be invariant under transformations $q\to 1-q$. We have verified numerically that this is indeed so (not shown).

\begin{figure}
\begin{center}
\includegraphics*[width=0.47\textwidth]{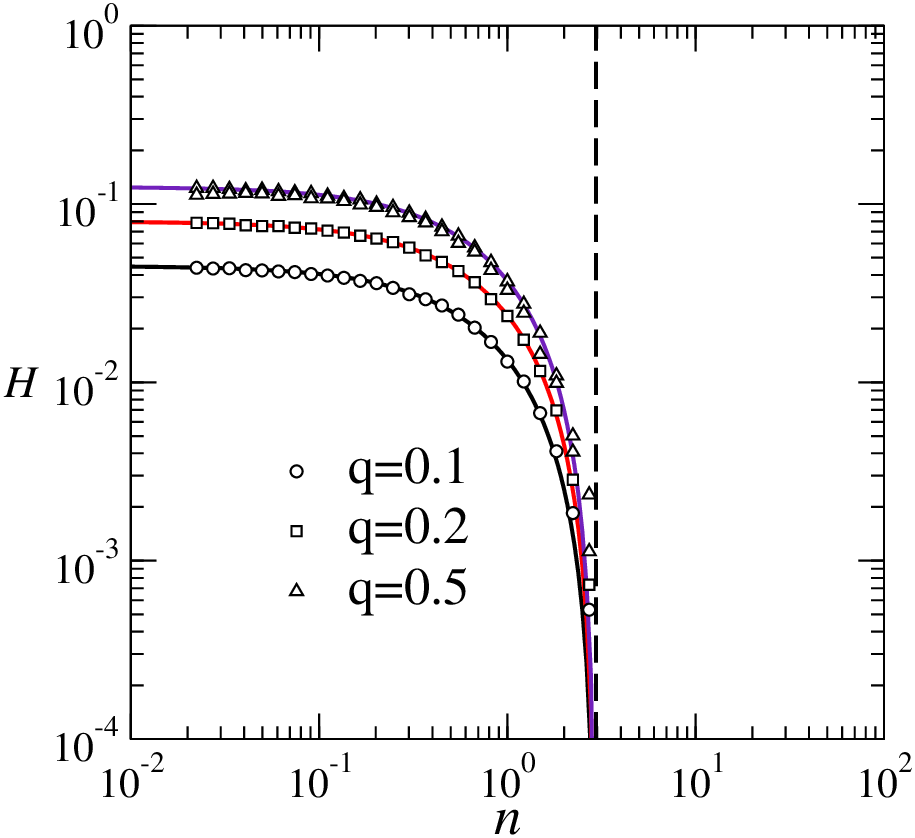}
\includegraphics*[width=0.45\textwidth]{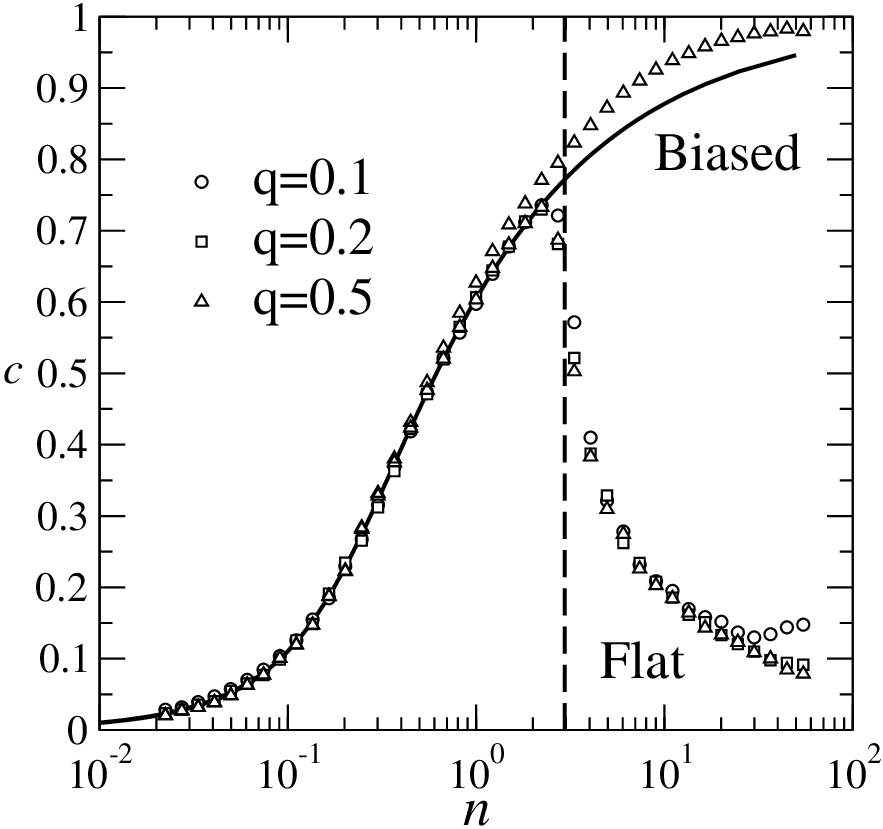}
\caption{\label{uno}Behavior of $H$ and $c$ versus $n$ for different values of $q$ and $S=2$. Markers correspond to results from computer simulations, averaged over 100 disorder samples. The dashed vertical line marks the position of the critical point $n_c$. The solid lines give the analytic result (valid only for $n<n_c$). ``Flat'' and ``Biased'' refer instead to different initial conditions: $p_i(0)=0$ and $p_i(0)=0.1$, respectively. For $H$ only results for $q=1/2$ are shown for different initial conditions, for simplicity.}
\end{center}
\end{figure}

Fig \ref{uno} shows the behavior of $H$ and $c$ as a function of $n$ for various $q$.
As expected, $H$ is similar to a physical order parameter. Its behavior indicates that as the number of consumers increases they tend to distribute more and more uniformly over producers until, for $n=n_c$ the distribution becomes uniform. For $n<n_c$ the economy is inefficient as the uneven distribution of demands generates exploitable profit opportunities. For $n>n_c$ the economy is instead efficient. Notice that results are indeed independent of initial conditions in the inefficient phase, while when $n<n_c$ the theory developed for the ergodic regime ceases to describe the steady state correctly. As a matter of fact, the steady state for ``flat'' initial conditions $p_i(0)=0$ for all $i$ and ``biased'' initial conditions $p_i(0)=0.1$ for all $i$ lead to very different regimes from a macroscopic viewpoint, as can be easily inferred from the behavior of $c$.

\begin{figure}
\begin{center}
\includegraphics*[width=0.6\textwidth]{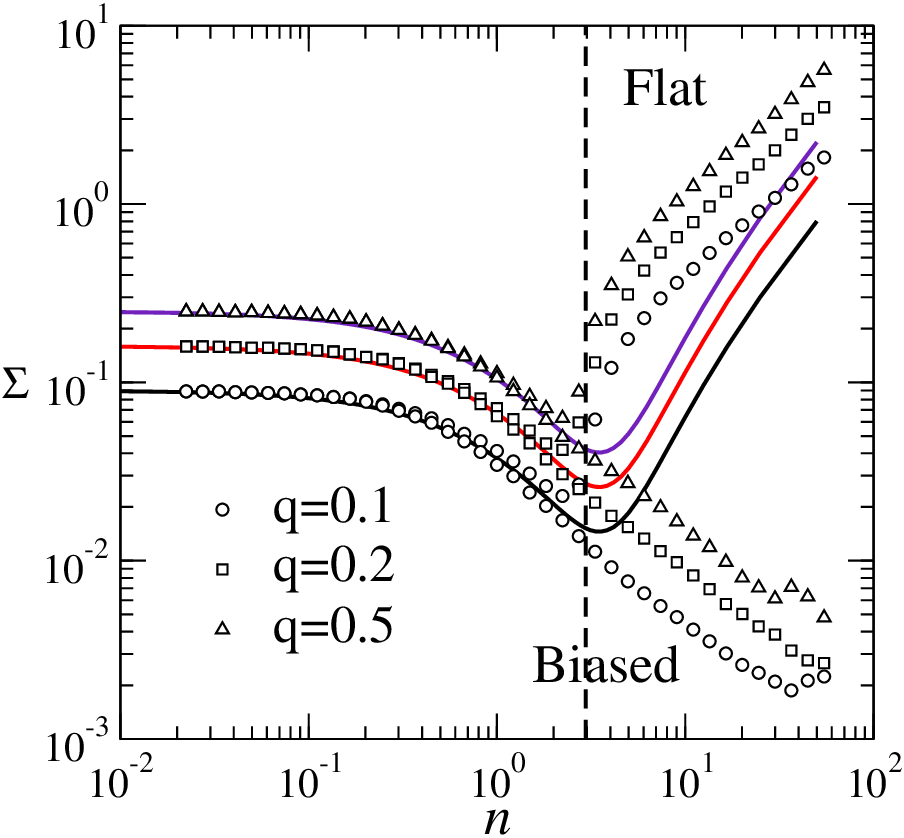}
\caption{\label{due}Behavior of $\Sigma$ (see text) versus $n$ for different values of $q$ and $S=2$. Markers correspond to results from computer simulations, averaged over 100 disorder samples. The dashed vertical line marks the position of the critical point $n_c$. The solid lines give the analytic result (valid only for $n<n_c$). ``Flat'' and ``Biased'' refer instead to different initial conditions: $p_i(0)=0$ and $p_i(0)=0.1$, respectively.}
\end{center}
\end{figure}

This is even more evident when one turns the attention toward fluctuations (see Fig. \ref{due}).
We show in particular the behavior of the quantity
\be
\Sigma=\frac{1}{P}\sum_\mu\avg{\l(D^\mu-\ovl{D}\r)^2}=\Delta+H
\ee
which also serves as a measure of the total amount of money invested by consumers. We see that in the inefficient phase fluctuations are small and well described by the approximate equations derived in the previous section. When the economy becomes efficient, however, the dependence on initial conditions may drive the system to both states with large price fluctuations where ($\Sigma\sim n$), which are rather undesirable, and states with small fluctuations (where $\Sigma\sim 1/n$). This can be interpreted with the following mechanism (borrowed from \cite{traf}). When there are few consumers, many sellers receive small demands and thus the economy presents many profitable opportunities. As more and more consumers join the opportunity window shrinks and players may be forced to switch bundles repeatedly in the attempt to identify convenient commodities. This leads to the increase of fluctuations and ultimately to a loss of day-by-day cost certainty. 

\begin{figure}
\begin{center}
\includegraphics*[width=0.47\textwidth]{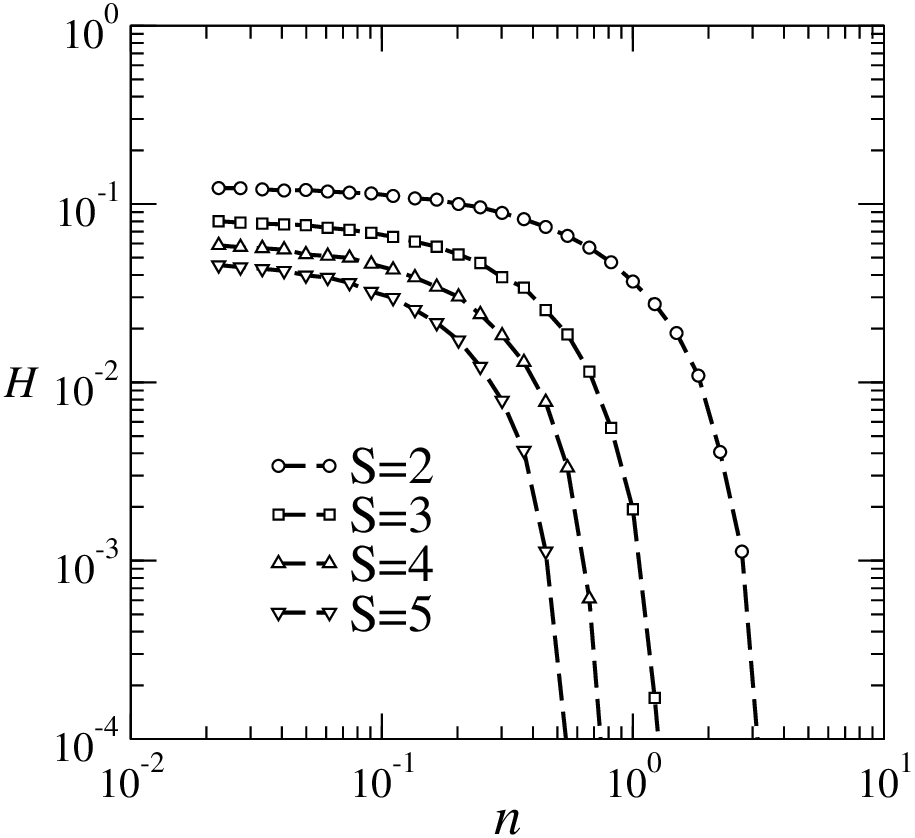}
\includegraphics*[width=0.45\textwidth]{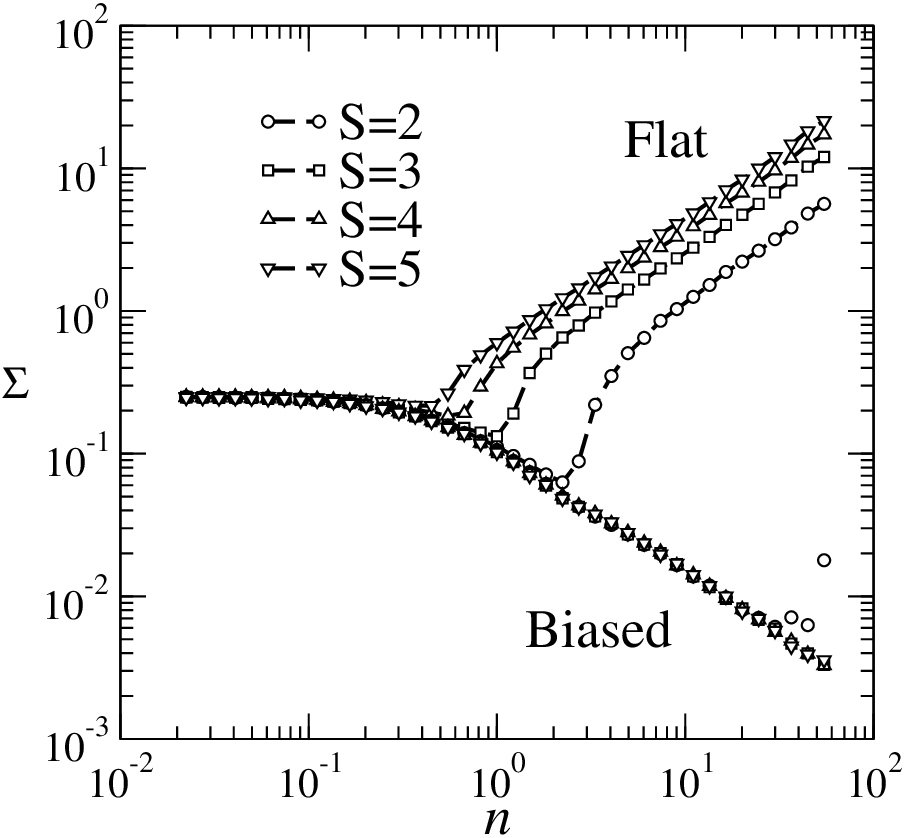}
\caption{\label{tre}Behavior of $H$ and $\Sigma$ versus $n$ for different values of $S$ and $q=1/2$. Markers correspond to results from computer simulations, averaged over 100 disorder samples. Lines are just a guide for the eye. ``Flat'' and ``Biased'' refer instead to different initial conditions: $p_i(0)=0$ and $p_i(0)=0.1$, respectively}
\end{center}
\end{figure}

Coming finally to the case of varying $S$ (for which we fix $q$ at 1/2, see Fig. \ref{tre}),
one sees that as the number of possible bundles per consumer grows, the phase transition is preserved although the critical point shifts to smaller values, indicating that a smaller number of consumer can fill the product space efficiently when consumers have more possible choices. However the peculiar behavior of fluctuations in the efficient regime is left unchanged.

\section{Conclusions}

The model presented here aims at studying how the choices made by adaptive consumers, and in particular how they distribute themselves over sellers, affect the resulting price distribution. We have seen that the interplay between the two aspects of the problem is rather subtle and definitely worth of further study. Many conclusions valid for Minority Games hold also for this model. In particular, it is possible to argue that if consumers may enter the market when there are many convenient opportunities and leave it when there is none, the economy would self-organize at the critical point $n_c$, which is the most efficient from a collective viewpoint. It is also interesting to notice that the regime with $H>0$ can be seen as one in which some consumers have identified a convenient seller from which they buy preferentially. Then the fact that $H>0$ can be interpreted as the formation of some sort of stable trading relationship (interestingly accompanied by a higher average price though smaller fluctuations). It may be possible to calculate $H$ and check whether the picture described here is realistic directly from empirical trading data from some specific markets in which agents trade frequently.  Examples that have received some attention in the economic literature are markets for perishable goods such as fish. Work along these lines is in progress. Finally, the model acquires yet more richness when consumers are given a degree of stochasticity in their decision making process, for example in the form of a finite learning rate, or when the dynamics of demand (which is the only side addressed here) is coupled to a supply dynamics generated by producers. In these cases, the behavior may depart significantly from that of the standard Minority Game (to some extent, this aspect was discussed in \cite{traf}). The extension of the present theory to that case will be reported elsewhere.

\acknowledgments

It is a pleasure to thank D Challet, R Mulet and G Weisbuch for useful comments and suggestions at different stages of this work.

\end{document}